\documentclass[12pt,preprint]{aastex}

\def\logz{\lbrack\hbox{M/H}\rbrack}

\newcommand{\mean}[1]{\langle #1 \rangle}

\slugcomment{To appear in the Astronomical Journal}
\shorttitle{Detection and Photometry of Clusters}
\shortauthors{Dolphin \& Kennicutt}

\begin{document}

\title{HST Survey of Clusters in Nearby Galaxies.\\
II.  Statistical Analysis of Cluster Populations}
\author{Andrew E. Dolphin}
\affil{Kitt Peak National Observatory, National Optical Astronomy Observatories, P.O. Box 26372, Tucson, AZ 85726}
\email{dolphin@noao.edu}

\and

\author{Robert C. Kennicutt, Jr.}
\affil{Steward Observatory, University of Arizona, Tucson, AZ 85721}
\email{robk@as.arizona.edu}

\begin{abstract}
We present a statistical system that can be used in the study of cluster populations.  The basis of our approach is the construction of synthetic cluster color-magnitude-radius diagrams (CMRDs), which we compare with the observed data using a maximum likelihood calculation.  This approach permits a relatively easy incorporation of incompleteness (a function of not only magnitude and color, but also radius), photometry errors and biases, and a variety of other complex effects into the calculation, instead of the more common procedure of attempting to correct for those effects.

We then apply this procedure to our NGC 3627 data from Paper I.  We find that we are able to successfully model the observed CMRD and constrain a number of parameters of the cluster population.  We measure a power law mass function slope of $\alpha = -1.50 \pm 0.07$, and a distribution of core radii centered at $r_c = 1.53 \pm 0.15$pc.  Although the extinction distribution is less constrained, we measured a value for the mean extinction consistent with that determined in Paper I from the Cepheids.
\end{abstract}


\section{Introduction}
As with stars, the star clusters of a galaxy provide direct fossil evidence of previous star formation events.  Compact clusters span the complete range of ages in the universe, from very recent star formation (30 Doradus) to the oldest components of galaxies (globular clusters).  Integrated photometry of a single cluster can provide an age and initial mass of that cluster, thus allowing one to trace the cluster formation history in a similar way that blue helium-burning stars trace the star formation history.  However, the effects of stellar evolution, cluster lifetimes, differential reddening, etc. prevent such a simple determination of physical parameters such as the cluster initial mass function and cluster formation rate.

Because of the difficulties in determining the physical parameters, work on cluster populations have used relatively crude statistical comparisons of observed properties.  The simplest is a measurement of the combined light of detected clusters, which can be compared with the brightness of the parent galaxy to determine a cluster formation rate parameter \citep{lar00}.  The advantage of such a technique is its relatively easy implementation and minimal amount of required data.  For example, WLM (a Local Group) galaxy has only one globular cluster.  One would not be able to determine a luminosity function, star formation history, or any other detailed parameters from this object; however a cluster frequency can be determined very accurately.

A more sophisticated technique involves an examination of the cluster luminosity function.  \citet{whi99} have made very effective use of this in their study of the Antennae, where there are many thousands of clusters to build up accurate statistics.  While a luminosity function contains considerably more information than does a cluster frequency, the complications listed above make the physical interpretation of the observed luminosity function extremely difficult, except under special conditions such as all clusters having the same age.

What is largely lacking is a comprehensive statistical technique to interpret the observational data in terms of physical properties.  Much progress has been made in the ability to determine star formation histories from stellar color-magnitude diagrams \citet{dol97,dol02b}; little progress has been made in the equivalent study of cluster populations.

This paper is the second in our series on a survey of the cluster populations of nearby galaxies.  Dolphin \& Kennicutt (2002; hereafter Paper I) demonstrated our procedure for the identification and photometry of clusters in our data, which consist of WFPC2 observations of galaxies between $\sim 4$ and $\sim 25$ Mpc.  In this paper, we are examining statistical methods of cluster populations with which we can constrain physical properties.

In section 2, we examine the many factors that affect the observed cluster population of a galaxy and describe a technique that will incorporate all of these to produce a synthetic cluster population for any set of parameters.  In section 3, we show how synthetic cluster populations and real cluster populations can be statistically compared to maximize the physical information; this technique is tested using simulated data in section 4.  Finally, we will conclude our study of NGC 3627 (begun in Paper I) with a demonstration of our statistical technique in Section 5.

\section{Synthetic Cluster Populations}

The key to accurate analysis of a galaxy's cluster population is an understanding of how various physical characteristics (such as cluster size) transform into observational quantities (such as sharpness).  Most observational quantities are functions of more than one physical characteristic.  For example a cluster's color is related to its age, metallicity, extinction, and the random selection of stellar masses; thus a simplistic attempt to measure one of those quantities using color is likely to fail.  In the same way, completeness is a function of color, magnitude, and cluster radius; an attempt to use the observed luminosity function to determine the cluster mass function is likewise inaccurate unless color and radius distributions are taken into account.  We will examine each of the physical parameters in this section.  Some will be seen to be very important; others will be less so.

Our analysis approach will center on generating and analyzing synthetic cluster populations.  Because our photometry provides three variables (magnitude, color, and sharpness), we will refer to color-magnitude-radius diagrams (CMRDs) instead of color-magnitude diagrams (CMDs).  In this section we describe the ingredients of a synthetic CMRD.  Our approach will be parallel to the star formation history measurement technique laid out by \citet{dol97,dol02b}; we will generate ``noiseless'' CMRDs that account for all possible combinations of age, metallicity, mass, and photometric effects in order to make statistical comparisons between these and the observed data.

\subsection{Cluster Formation History}

The primary influences on a galaxy's cluster population are its star formation history (mass of stars formed as a function of age and metallicity), cluster formation efficiency (mass of clusters formed divided by mass of stars formed), and chemical enrichment history.  However, constraints for these quantities are very poorly-determined.  The present-day mean metallicity is $\logz = +0.2 \pm 0.2$ (Kennicutt et al., in prep) from nebular spectroscopy.  We can reasonably assume that the mean metallicity has slowly and consistently increased from metal-poor globular abundances (for which we use $\logz = -2.3$, the most metal-poor of the \citet{gir00} isochrones) at ancient ages to the present-day value, which allows us to estimate an age-metallicity relation.  We note that the metallicities at older ages are not as important, since clusters dim so rapidly that most of what we observe is younger than $20-40$ Myr.  Since there is some scatter in the metallicities of HII regions within a galaxy, we must also allow for scatter around our assumed age-metallicity relation.  We adopt an rms scatter of $\pm 0.13$ dex, based on the Galactic disk study of \citet{roc00}.

The star formation rate and cluster formation efficiency, however, are nearly completely unconstrained from the $V$ and $I$ imaging alone.  However, the timescale over which clusters are visible is extremely brief.  Our NGC 3627 data (Paper I, figure 15) shows that even a cluster with initial mass of $10^5 M_{\odot}$ would fade below the bottom of our observed CMRD in $20-40$ Myr.  We therefore will make the assumption that the cluster formation rate (the product of the star formation rate and cluster formation efficiency) has been constant over the observable period.  This assumption would only be invalid in the case where a short burst created a large number of clusters in the same evolutionary phase; such a situation would be obvious from an examination of the CMRD, however, as all clusters formed in a burst will have roughly the same color.  We recognize that, during the epoch of globular cluster formation, the cluster formation rate was likely much different than what is observed presently.  Thus our synthetic cluster CMRD will consist of two epochs of cluster formation, with a dividing line of 10 Gyr.  These two populations will be distinct in broadband $VI$ photometry, as the old population consists entirely of red ($V-I > 0.8$) clusters while the younger population contains a mixture of colors.

In the case of NGC 3627 (analyzed below in Section 5), we indeed find that the form of our cluster formation rate is unimportant, and that the 2-parameter star formation rate fits the data as well as a more sophisticated function.  We also created synthetic CMRDs using a variety of metallicity enrichment laws (with the same current metallicity), all of which produced similarly-good fits to the data.

We therefore characterize the cluster formation history with four parameters: cluster formation rates at young and ancient ages, an age-metallicity relation, and scatter around the age-metallicity relation.  However, while all of these ingredients are necessary to produce a synthetic CMRD, the only parameter that significantly affects the CMRD is the recent cluster formation rate.

\subsection{Initial Mass and Radius Distributions}

Once we have determined (or assumed) the cluster formation rates and the age-metallicity relation, we know the mass of clusters formed at each age and metallicity.  To produce a synthetic CMRD, however, we need the actual number of clusters formed.

We first divide the mass of clusters at one age and metallicity according to an assumed cluster initial mass function.  The function can be anything; we will assume a power law of the form
\begin{equation}
\frac{d \log ( dN / d \log M )}{d \log M} = \alpha,
\end{equation}
or
\begin{equation}
\frac{d N}{d \log M} \propto M ^{\alpha}.
\end{equation}
As this function diverges at low mass for $\alpha \le -1$, we arbitrarily choose a low-mass cutoff of 1 solar mass for now, and we hope that we can constrain the cluster initial mass function at low mass from the closer galaxies in our sample.

The second division of clusters will be according to cluster radius.  This is extremely difficult, as our observed data have very strong selection effects in radius: small clusters will be classified as stars while large clusters will be too difficult to photometer or may be identified as small associations and thus rejected.  Because we are sensitive to only a limited range of core radii ($0.2 - 4$ pc in our NGC 3627 data), we will initially assume a flat distribution of cluster radii,
\begin{equation}
\frac{d N}{d r} = constant.
\end{equation}
We will be able to constrain this further by comparing the sharpness distributions of the observed and synthetic CMRDs.

With the clusters now divided by age, metallicity, initial cluster mass, and core radius, one other factor affects the physical properties of the cluster population: lifetimes.  However, in terms of generating a synthetic CMRD, we consider this effect unimportant due to the very short timescales over which clusters are visible above our photometric cutoffs.  With our limiting magnitude of $V = 24$, we only require that a cluster with an initial mass of $10^4 M_{\odot}$ lives $10-15$ Myr and that one with a mass of $10^5 M_{\odot}$ lives $30-40$ Myr.  (That the upper end of the mass function is unaffected by disruption is demonstrated by the analytical models of Fall \& Zhang 2001.)  As cluster lifetimes are expected to be at least an order of magnitude beyond these values \citep{kro01b,fal01}, we can thus ignore the effect of cluster disruption.

We have thus added two additional parameters to those listed at the end of the previous section: the cluster initial mass function and radius distribution.

\subsection{Intrinsic Observed Properties}

Accounting for the factors listed in the previous two sections, we now know the physical properties (age, metallicity, mass, and core radius) of the cluster system.  These must now be converted into the intrinsic observable quantities of magnitude, color, and angular size.  The final quantity is trivially determined from the core radius and distance; we address the others in this section.

The magnitude and color of a cluster are generally determined using stellar evolution models \citep{bru93,gir00}.  We will adopt that approach here as well, using the models of \citet{gir00} and the revised stellar IMF of \citet{kro01} to determine integrated magnitudes and colors.  We acknowledge that the theoretical stellar models are not perfect; however we believe that they are sufficiently accurate that integrated magnitudes of clusters should be reasonably correct, even if a few evolutionary phases are modeled inaccurately.  At any rate, as no empirical database of integrated cluster magnitudes exists spanning the entire range of ages, metallicities, and filter passbands that we will need, theoretical models must be used.

There is an additional complexity that we must account for.  Because the division of a certain amount of mass into stars does not always produce the same set of star masses, there is scatter inherent in the cluster colors.  In other words, clusters of the same age and metallicity do not have exactly the same color and magnitude.  We therefore used Monte Carlo simulations to determine the color and magnitude distribution of clusters with a wide range of age, metallicity, and initial mass; the full distribution is added to our CMRD rather than merely the mean magnitudes and colors.

For example, the mean cluster of solar metallicity, age 10 Myr, and initial mass of $10^5 M_{\odot}$ has magnitudes of $M_V = -11.89$ and $V-I = 0.71$.  Because of noise in the distribution of stars, however, this population of clusters appears as a ``smear'' with $\sigma_V = 0.08$ magnitudes and $\sigma_{V-I} = 0.07$ magnitudes.  This factor becomes extremely important for young clusters with lower masses, in which case the scatter from the distribution of stars becomes much larger than the photometric errors.

Once the distribution of magnitudes and colors is determined, we adjust the magnitudes for distance and global extinction.  For NGC 3627, we determined the distance from Cepheids and adopted the foreground extinction of $A_V = 0.11$, calculated from the maps of \citet{sch98}.  Because most (if not all) the galaxies in our sample have ongoing star formation and thus dust, we must also account for differential extinction.  We therefore apply a range of extinction values, requiring the lowest value to equal the foreground extinction and the mean value to equal the extinction measured from the Cepheids.  We will begin by assuming a flat distribution of extinction values; this can be constrained slightly by the data.

\subsection{Observational Errors and Selection Effects}

The final ingredient needed to generate a synthetic CMRD is the artificial cluster tests.  These were described in full in Paper I; we will summarize here.  Our artificial cluster tests are identical in concept to artificial star tests: we add artificial clusters to the image (using an assumed cluster profile convolved with the stellar PSF), photometer them, and apply the same magnitude, sharpness, and roundness cuts that were applied to the real data.  This produces an artificial cluster library that characterizes not only the completeness fraction, but also the recovered magnitude, color, and sharpness as a function of input magnitude, color, and angular size.

Using this library, we can thus reproduce the photometry of the galaxy being studied, as well as \textit{exactly} reproduce our selection effects.  This is the reason why we chose very conservative magnitude and sharpness limits in Paper I; accounting for false positive detections is much more difficult than simulating selection effects.

\subsection{Building a Synthetic CMRD}
\label{sec_cmrdsumm}

With all of the ingredients described in this section, we are able to generate a synthetic cluster CMRD.  Summarizing the parameters listed above, the following pieces of information are all that is required:
\begin{enumerate}
\item Young and ancient cluster formation rates
\item Age-metallicity relation and scatter
\item Cluster mass and radius distributions
\item Distance, extinction, and differential extinction
\item Synthetic models from isochrones, including intrinsic scatter in colors and magnitudes
\item Artificial cluster library
\end{enumerate}
Before we begin an analysis of a galaxy, the final two items are already calculated.  The synthetic models are pre-computed, while the artificial cluster library is part of our photometry process.  The distance is measured from Cepheid photometry, as is an initial guess of the extinction.  We assume an age-metallicity relation, and assume functional forms for the cluster initial mass function, radius distribution, and differential extinction.  An initial guess as to the mass function can be taken from the luminosity function.  As noted above, the resulting CMRD is almost completely insensitive to the age-metallicity relation, provided that the current metallicity is fixed to the observed value.  This leaves only two completely unknown parameters (the young and ancient cluster formation rates) and three constrainable parameters with initial guesses (extinction, mass function, and radius distribution).

\section{Cluster Population Statistics}

The previous section describes how we generate a noiseless synthetic CMRD, which samples all possible ages, metallicities, masses, radii, extinction values, and photometric effects.  As described in section \ref{sec_cmrdsumm}, all but three parameters are known or assumed before we begin our analysis: the young cluster formation rate, the ancient ($> 10$ Gyr) cluster formation rate, and the slope of the cluster initial mass function.  These three values can be determined by a statistical comparison of the observed and synthetic CMRDs.

Computationally, we characterize each CMRD as a three-dimensional matrix, with the dimensions corresponding to magnitude, color, and sharpness.  Within each element of the matrix is the number of observed or synthetic stars with that magnitude, color, and sharpness.  Because of the nature of WFPC2 data, various parts of the image will have different statistical characteristics.  For NGC 3627, for example, WFC2 contained severe crowding (because of its proximity to the center of the galaxy) not seen in the other chips.  Additionally, the PC always has different characteristics than the other chips because of its different plate scale.  Thus we produce multiple CMRDs, one corresponding to each region.  For most galaxies, we will use one CMRD per chip.  When studying galaxies with photometry in three or more bands, we will again multiply the number of CMRDs to allow the simultaneous examination of, for example, a (B, B-V) CMRD and a (V, V-I) CMRD.

We judge the level of agreement between the observed and synthetic CMRDs by the probability that the observed data were drawn from the synthetic CMRD.  Because our synthetic CMRD is noiseless (rather than randomly populated), we can use the Poisson probability law:
\begin{equation}
P = \prod_i \frac {e^{-m_i} m_i^{n_i}} {n_i !},
\end{equation}
where $P$ is the probability, $i$ represents the CMRD element, $m$ is the synthetic (model) CMRD, and $n$ is the observed CMRD.  Maximizing this value for an unchanging observed CMRD is the equivalent of minimizing
\begin{equation}
\label{eq_fitparam}
2 \sum_i m_i - n_i + n_i \ln \frac{n_i}{m_i},
\end{equation}
the Poisson equivalent of $\chi^2$.  By determining the combination of cluster formation rates and mass function slope that minimizes this equation, as well as the acceptable range of values surrounding the minimum, we can thus determine these values empirically.  We also anticipate being able to constrain the cluster radius distribution.  Once we have compiled sufficient data, we will also attempt to constrain the lower-mass cutoff (or any slope change) in the mass function.

Details of the statistics, including a derivation of this formula, an explanation of why $\chi^2$ cannot be used with Poisson-distributed data, importance of using noiseless synthetic CMRDs, and techniques for evaluating the quality of the best fit and uncertainties in the parameters are found in \citet{dol02b}; we do not wish to duplicate the lengthy discussion of statistics here and instead refer the reader to that paper.

\section{Synthetic Galaxy}

Before concluding our example analysis of NGC 3627, we wish to demonstrate the effectiveness of our analysis technique via application to a simulated galaxy.  To make the results of this exercise directly comparable to our NGC 3627 study, we will use the parameters producing the best fit to NGC 3627's cluster population.  These values are shown as the ``input'' values in Table \ref{tab_synthfit}.
\placetable{tab_synthfit}

\subsection{Fit Accuracy}

Our first test was to determine the parameters that produced the best fit to the synthetic data.  A comparison between these values and those used to create the synthetic data will test our technique for any biases.  While some difference is expected (because the simulated observed CMRD was randomly populated), one expects that the differences between the recovered and input values will be less than the uncertainties.  Indeed, this was true, as is seen in Table \ref{tab_synthfit}.  Additionally, the recovered cluster formation rate differed from the input rate by less than 1\%.  Thus, with all values being measured within their uncertainties, we have no reason to suspect biases in the method.

\subsection{Correlated Parameters}

Our second test is a search for correlated errors in the parameters, which was done by forcing each of the parameters, in turn, to a ``bad'' value ($1 \sigma$ from the best value) and looking for any effects on the best fits of the other parameters.  Our expectation is that the mean value and width of the radius distribution will be correlated with each other (since both are primarily related to the cluster sharpness), but that the cluster IMF slope (primarily a function of $V$ magnitude) and extinction distribution (primarily a function of $V-I$ color) will be independent.  Table \ref{tab_synthcorr} shows the results of this test, with deviations given in units of $\sigma$.  One can easily discern the correlations between the two radius parameters, as well as their independence from the other two parameters.  What is more surprising is the apparent correlation between the cluster IMF slope and the extinction distribution.  Since no such correlation should exist (a pure power-law luminosity function will keep the same slope, regardless of extinction), we assume this to be a numerical instability created by the poorly-constrained nature of the extinction, rather than a true correlation.  We note that the change in the mass function slope created by the extinction change is extremely small ($0.05$); it appears as a large number in the table because the uncertainty in the measured value is $\pm 0.07$.
\placetable{tab_synthcorr}

\section{NGC 3627}

With the mechanism now in place for analyzing the cluster population, we return to the study of NGC 3627 (M66) begun in Paper I.  As noted in Paper I, this galaxy is part of an interacting system.  We found 528 clusters, which account for $\sim 1$\% of the galaxy's luminosity.  By re-reducing the data with a newer version (1.1) of HSTphot \citep{dol00}, the number of clusters is now 506.  We have also conducted extensive artificial cluster tests, in which clusters of known integrated magnitudes and radii are inserted and photometered; this allows us to accurately model observational effects such as incompleteness, photon noise, crowding, and photometric biases.

For our analysis, we have reduced the data to two sets.  First is the set of 506 observed clusters, for which we need only the observed $V$ and $I$ magnitudes, as well as the sharpness measurement.  Photometry uncertainties are not needed, as these are simulated using the artificial clusters.  The second data set is the artificial cluster library, which contains input integrated $V$ and $I$ magnitudes and core radii, as well as the observed properties (or a flag if the cluster was either not recovered or did not meet our cluster selection criteria).  The CMRD modeling then proceeded according to the steps described in the previous sections.

\subsection{Fit Quality}

Although this modeling is not quite as sophisticated as the stellar population CMD-fitting method of \citet{dol02} due to a number of approximations in our synthetic CMRD construction, we believe that it is useful to look at the fit quality to determine how well we matched the observed data.  The maximum likelihood fit parameter (equation \ref{eq_fitparam}) of our best solution was 924.7.  For comparison, the typical data set randomly drawn from our best model would produce a fit parameter of 850.9, with a $1 \sigma$ scatter of 38.2.  Thus the match between the observed data and our best model is roughly 2 $\sigma$ from ideal.  Given the approximations involved in using functional forms for differential extinction and the mass and radius distributions, we consider this a good fit.

\begin{figure}
\plotone{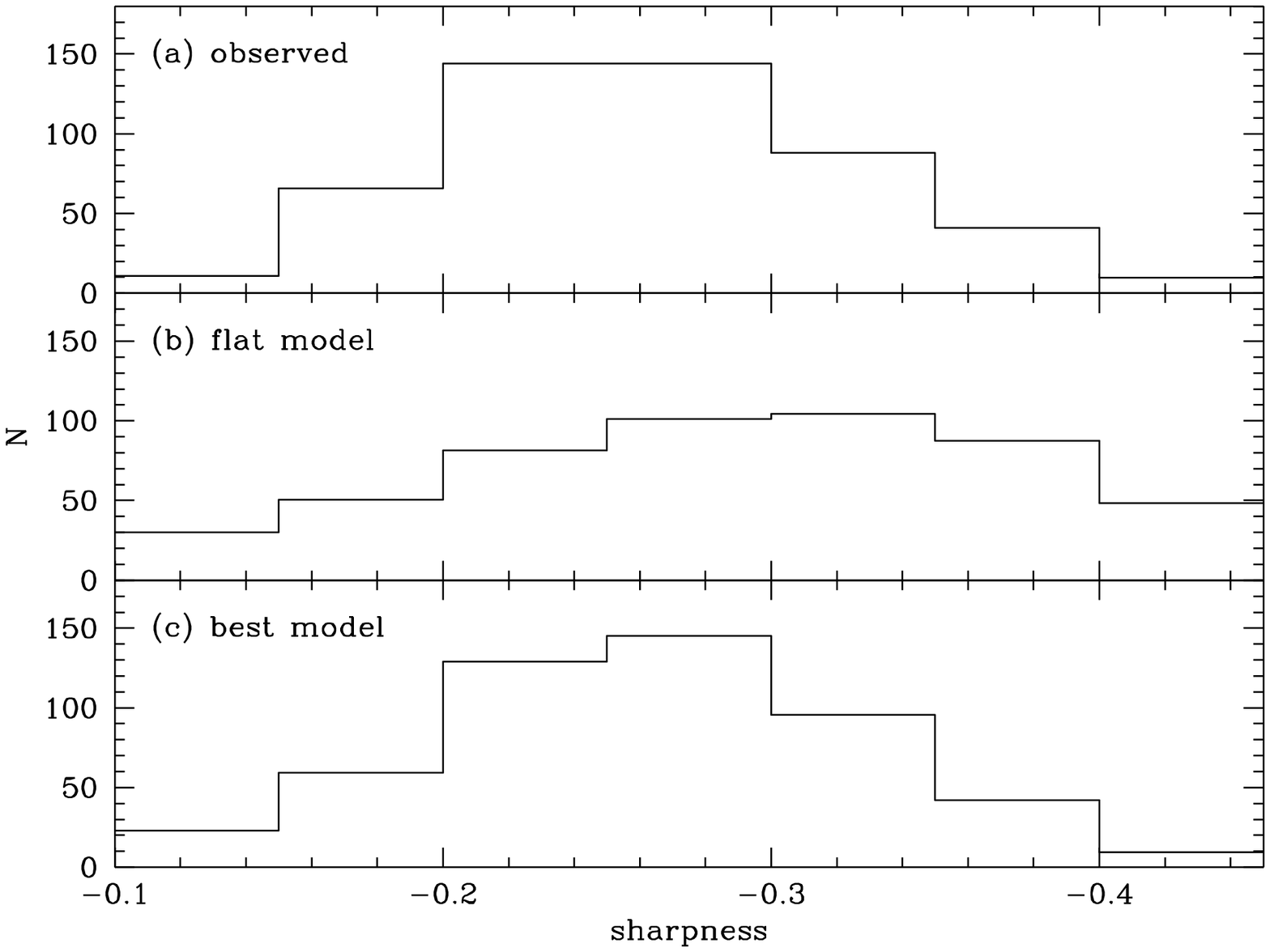}
\caption{Observed and simulated distributions of sharpness.  The top panel shows the histogram of sharpness values seen in the observed data.  The middle and bottom panels show the distributions measured from two synthetic CMRDs; the middle panel was created using a flat distribution of core radii, while the bottom panel is our best-fitting model.  The excellent agreement between the top and bottom panels is expected; however a discrepancy would have indicated of a problem with our analysis technique.}
\label{fig_sharp_dist}
\end{figure}

\begin{figure}
\plotone{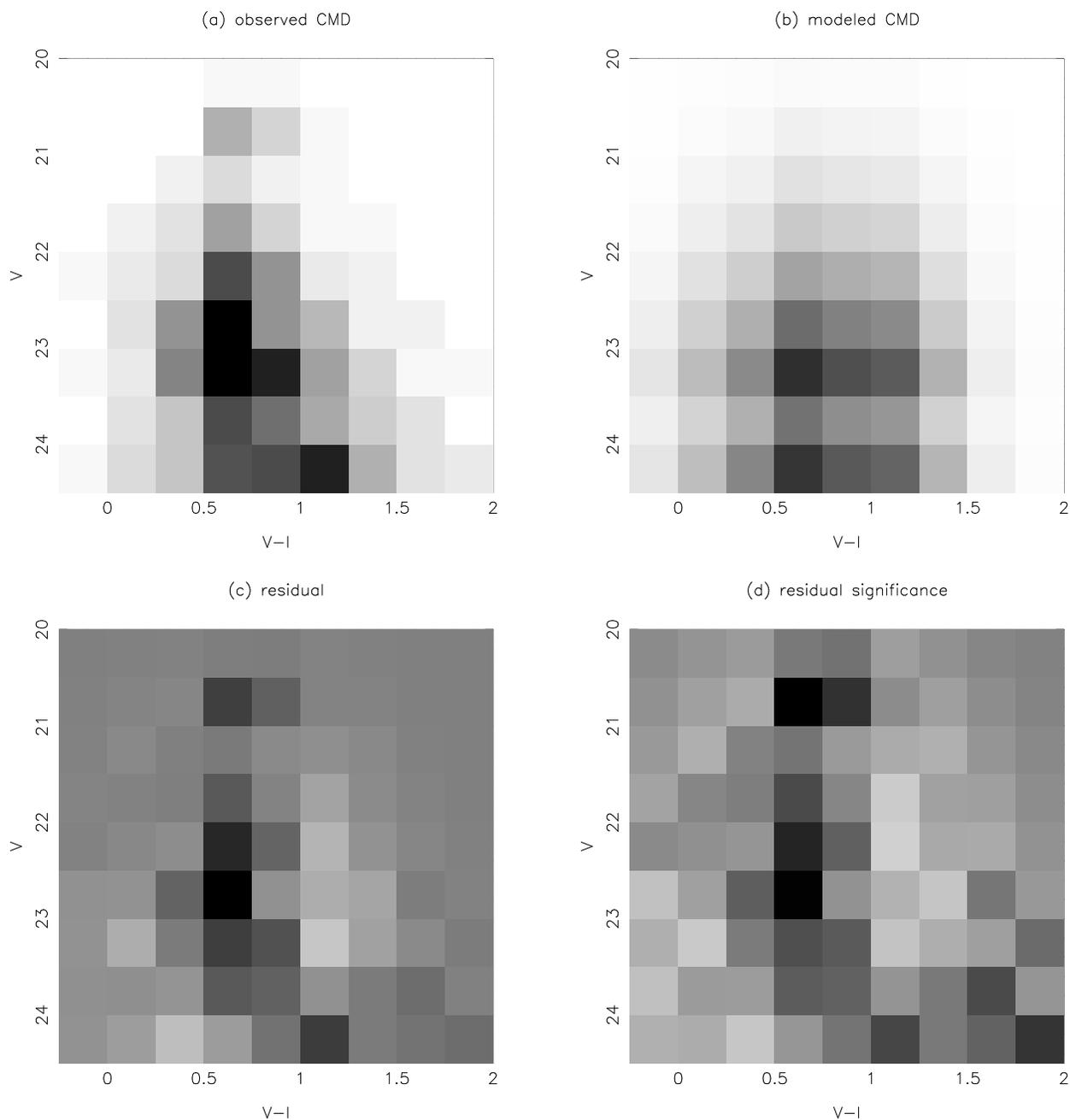}
\caption{Greyscale color-magnitude diagrams.  Upper left and upper right show the observed and synthetic CMDs, respectively.  The bottom left diagram is the residual, while the bottom right shows the significance of the residuals.  The apparent break at $V = 23.5$ is caused by the higher magnitude restriction imposed on Chip 2, as described in Paper I.  The greyscale in panels (a), (b), and (c) has the same stretch; that in panel (d) stretches from $-3 \sigma$ to $+3 \sigma$.}
\label{fig_cmrds}
\end{figure}

A comparison between the top and bottom panels of Figure \ref{fig_sharp_dist} shows that our best model matched the observed sharpness distribution very well.  A comparison of the CMDs is given in Figure \ref{fig_cmrds}.  The top panels show the observed (left) and model (right) CMDs in greyscale; the break at V=23.5 is caused by the higher cutoff magnitude in WFC2 that was imposed to combat contamination.  (Since the magnitude limits were also applied to the synthetic data, it is possible to include photometry of varying depth without any extra effort.)  The bottom left panel shows the difference of the top panels, with darker squares signifying more observed than synthetic clusters and lighter squares the reverse.  The important panel of Figure \ref{fig_cmrds} is the bottom right panel, which shows the significance of the differences (essentially a $\chi$ measurement, though using Poisson statistics).  In no part of the CMD is that fit worse than $3 \sigma$, and only two CMD regions are worse than $2 \sigma$.  (Note that the CMRD as fit by our algorithm is truly a three-dimensional diagram; we have divided it into a CMD and a radius histogram in this section merely for ease of display.)
\placefigure{fig_sharp_dist}
\placefigure{fig_cmrds}

\subsection{Cluster Initial Masses}

Of the free parameters that enter into the solution, two are quite well-constrained.  The first is the cluster initial mass function slope.  Because the luminosity function of a population of coeval clusters is nearly the same as its mass function, and because the sum of many power laws with identical slopes is another power law with that slope, the luminosity function of a cluster population is a power law with the same slope as its mass function.  Two factors complicate this slightly.  First, the random division of a cluster's mass into stars will cause the mass-to-light ratios of coeval clusters to be slightly different.  As the scatter is larger for low-mass clusters, and such clusters are more common, the luminosity function is systematically flattened slightly.  The second complication comes from incompleteness, which is a function of magnitude, color, and cluster size.  Since completeness is lower at fainter magnitudes, this again flattens the luminosity function.  Both of these are second-order effects, however; the relationship between the mass function and luminosity function is very strong and thus it is the best-constrained of the free parameters.

Using a trial-and-error approach (creating a variety of model CMRDs and determining which fits the best), we measure a mass function slope of $\alpha = -1.50 \pm 0.07$.  This is slightly steeper than the observed luminosity function slopes of $-1.43 \pm 0.12$ (all clusters) and $-1.22 \pm 0.23$ (blue clusters) calculated in Paper I; this was expected since those values did not incorporate completeness corrections.

\subsection{Cluster Radii}

\begin{figure}
\plotone{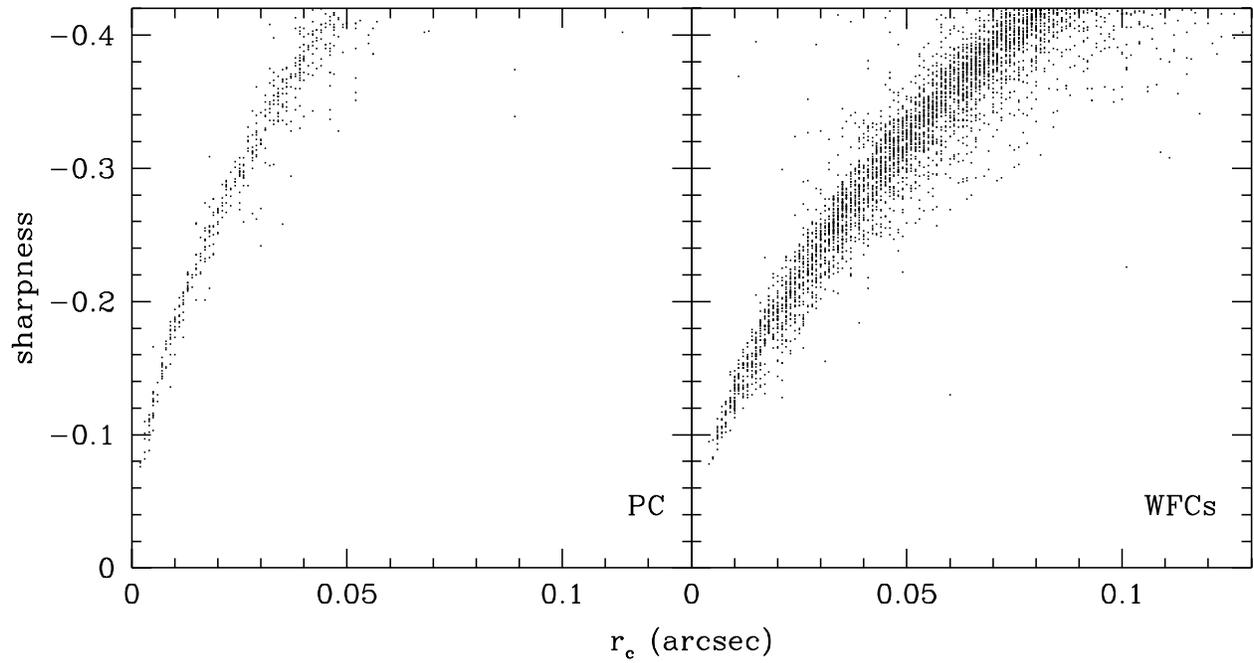}
\caption{Measured sharpness vs. input core radius for artificial clusters.  The left panel shows clusters in the PC; the right panel shows clusters in the WFCs.}
\label{fig_sharp_rc}
\end{figure}

The second well-constrained input parameter is the distribution of cluster radii.  As shown in Figure \ref{fig_sharp_rc}, the sharpness value measured by HSTphot is an extremely accurate measurement of the cluster's radius.  While there is some scatter in the diagram, the relation is tight in all chips, with sharpness proportional to the square root of the radius.
\placefigure{fig_sharp_rc}

While the scatter in the sharpness vs. core radius relation and the strong influence of radius on completeness prevent one from determining the intrinsic radius distribution from the CMRD in the way that one can determine a luminosity function, it is not difficult to constrain the radius distribution by comparing the observed and synthetic CMRDs.  Figure \ref{fig_sharp_dist} shows the distribution of observed sharpness values, as well as those created by two different models.  Because the middle panel (computed using a flat distribution of core radii) fits the observed data poorly, we can conclude that the shape of the observed distribution is not solely caused by selection effects and incompleteness.  Instead, we are able to solve for the radius distribution.

The best-fitting sharpness distribution is shown in the bottom panel of Figure \ref{fig_sharp_dist}, and was created with a Lorentzian distribution of core radii centered at $r_c = 1.53 \pm 0.15$ pc and with a HWHM of 0.88 pc.  The mean core radius is consistent with the results of \citet{els87}, who studied the profiles of 10 young LMC clusters.  The mean core radius of their sample of clusters was 2.19 parsecs, though the distribution is skewed by a few large clusters.  The peak of their distribution is near a core radius of 1.7 parsecs with the values for most clusters falling between 1.3 and 2.

\subsection{Extinction}

Unlike the distributions of cluster mass and radii, the extinction profile is not readily apparent from the CMRD.  We therefore took our initial guess of the extinction based on three observable values.  The lower extinction limit was set, based on the maps of \citet{sch98}, at $A_V = 0.11$.  The upper limit was determined by our reddest clusters; with no bright clusters redder than $(V-I) = 1.8$, there was no need to include extinction values greater than $A_V = 2.0$.  The final initial constraint was our observed mean Cepheid extinction values of $0.58 \pm 0.12$ using the \citet{mad91} period-luminosity relation and $0.76 \pm 0.11$ using the \citet{uda99} relation.

Using these loose constraints, we attempted to fit the data using a variety of distributions.  Because of the weak relationship between the extinction distribution and the shape of the CMRD, this solution was very poorly constrained.  (The extinction determination was constrained by the color distribution of the clusters, but cluster formation history also affects color distribution and therefore the two factors are partially degenerate.  Note that many of the galaxies we plan to study do have images in three broadband filters; in these cases the degeneracy can be more easily broken.)  The function producing the CMRD best matching the data was a power law of the form
\begin{equation}
\frac{dN(A_V)}{dA_V} = A_V^{-0.85 \pm 0.36}.
\end{equation}
Within our limiting extinction values, this distribution has a mean extinction of $\mean{A_V} = 0.71 \pm 0.15$, which is in agreement with both measurements from the Cepheids.

\subsection{Cluster Formation Rates}

The final parameters in the solution are the young and ancient cluster formation rates.  Unfortunately, the values we calculate are of limited value for a variety of reasons.  Most importantly, from a single set of deep WFPC2 images, we have no reliable constraint on the mass function at high (due to WFPC2 saturation) or low masses (due to a loss of signal).  Thus, while we were able to accurately measure the slope and scale of the mass function in the range where these data are sensitive, any estimate of the total mass contained in clusters is uncertain by many orders of magnitude.  Note that calculating the combined cluster mass for the range of masses to which we are sensitive is not productive, since our galaxy sample includes objects spread over 5 magnitudes in distance modulus and WFPC2's dynamic range (decreased by the high background under clusters) is only about 5 magnitudes.  Therefore the set of clusters we can see at $25-30$ Mpc do not overlap from those we can see at $2.5-3$ Mpc.

Our future papers will solve this problem in two ways.  The direct approach is to observe galaxies at larger and smaller distances; examination of their luminosity functions will allow us to better-understand the cluster mass function at high and low mass.  Even should that not provide adequate constraints, we can accurately determine \textit{relative} cluster formation rates of a sample of galaxies by using the same mass function for all.

Secondly, our functional form of the extinction distribution is almost certainly an approximation.  Since both the extinction and the cluster formation history affect the colors, any error in the measured extinction law will affect the derived cluster formation rates.  Thus we do not feel that the cluster formation history of any one system can be accurately measured unless a third band of photometry is obtained.

\section{Summary}

We have presented a statistical system that can be used to study cluster populations.  Our approach is to build synthetic cluster CMRDs for a variety of cluster population parameters, rather than attempting to take measurements directly from the observed data.  In short, we generate a set of synthetic CMRDs based on a variety of cluster population properties and use a Poisson likelihood statistic to determine which is the best.  Because of a variety of factors (completeness that varies as a function of color, magnitude, and radius; biases in the photometry; Poisson noise in the stellar composition of same-mass clusters; etc.), we believe that it is impossible to make definitive claims based on an examination of only the observed CMD; this statistical approach provides the capability of accounting for all such factors.

This approach is tested on the NGC 3627 data set from Paper I.  We found that we were able to create a model that matched the observed data very well, both in reproducing the observed distribution of sharpness and the observed CMD.

The cluster mass function and radius distributions were very well constrained from the data.  The cluster mass function is primarily determined by the cluster magnitudes, while the radius distribution is constrained by the sharpness measurements.  Thus we can accurately determine a mass function slope of $\alpha = -1.50 \pm 0.07$ and a distribution of core radii centered on $r_c = 1.53 \pm 0.15$pc.  Less certain were our measurements of the extinction distribution and cluster formation history, as both affect the colors.  These variables will be more strongly constrainted in our future work, which will include galaxies observed through three broadband filters.  Nevertheless, we were able to measure a mean extinction that was consistent with that measured from the Cepheids, giving some indication that even that poorly-constrained parameter was indeed measured correctly.

\acknowledgments

Support for this work was provided by NASA through grant number AR-09196.01 from the Space Telescope Science Institute.  All of the data presented in this paper were obtained from the Multimission Archive at the Space Telescope Science Institute (MAST). STScI is operated by the Association of Universities for Research in Astronomy, Inc., under NASA contract NAS5-26555.

\clearpage

\clearpage

\clearpage

\clearpage
\begin{deluxetable}{lrr}
\tablecaption{Input and recovered parameters of a synthetic cluster system. \label{tab_synthfit}}
\tablewidth{0pt}
\tablehead{
\colhead{value} &
\colhead{input} &
\colhead{recovered}}
\startdata
IMF slope    & $-1.50$   & $-1.50 \pm 0.07$ \\
$\mean{r_c}$ & $1.53$ pc & $1.43 \pm 0.13$ pc \\
$r_c$ HWHM   & $0.88$ pc & $0.86 \pm 0.16$ pc \\
$A_V$ slope  & $-0.85$   & $-0.76 \pm 0.38$ \\
\enddata
\end{deluxetable}

\clearpage
\begin{deluxetable}{lrrrrr}
\tablecaption{Correlations between the four fit parameters. In each test, the altered value was changed by $+1 \sigma$, and changes in the best fits of the other three values measured in units of $\sigma$. \label{tab_synthcorr}}
\tablewidth{0pt}
\tablehead{
\colhead{altered value} &
\colhead{IMF slope} &
\colhead{$\mean{r_c}$} &
\colhead{$r_c$ HWHM} &
\colhead{$A_V$ slope}}
\startdata
IMF slope    & $. . .$ & $-0.05$ & $+0.03$ & $-1.26$ \\
$\mean{r_c}$ & $-0.03$ & $. . .$ & $-0.24$ & $+0.00$ \\
$r_c$ HWHM   & $+0.01$ & $-0.42$ & $. . .$ & $+0.00$ \\
$A_V$ slope  & $-0.68$ & $+0.01$ & $+0.00$ & $. . .$ \\
\enddata
\end{deluxetable}

\end{document}